\documentclass[
]{ceurart}

\usepackage{listings}
\begin{document}

\copyrightyear{2024}
\copyrightclause{Copyright for this paper by its authors.
  Use permitted under Creative Commons License Attribution 4.0
  International (CC BY 4.0).}

\conference{PhysioCHI: Towards Best Practices for Integrating Physiological Signals in HCI,   May 11, 2024, Honolulu, HI, USA}

\title{Integrating Physiological Data with Large Language Models for Empathic Human-AI Interaction}

\tnotemark[1]
\tnotetext[1]{You can use this document as the template for preparing your
  publication. We recommend using the latest version of the ceurart style.}

\author[1, 2]{Poorvesh Dongre}[%
email=poorvesh@vt.edu,]
\cormark[1]
\address[1]{Department of Computer Science, Virginia Tech, Blacksburg, VA, 24060}
\address[2]{Empathic Computing Lab, Auckland Bioengineering Insitute, University of Auckland, New Zealand}

\author[1]{Majid Behravan}[%
email=behravan@vt.edu,
]

\author[2]{Kunal Gupta}[%
email=kgup421@aucklanduni.ac.nz,
]

\author[2]{Mark Billinghurst}[%
email=mark.billinghurst@auckland.ac.nz,
]

\author[1]{Denis Gra{\v{c}}anin}[%
email=gracanin@vt.edu,
]

\cortext[1]{Corresponding author.}
\fntext[1]{These authors contributed equally.}

\begin{abstract}
 This paper explores enhancing empathy in Large Language Models (LLMs) by integrating them with physiological data. We propose a physiological computing approach that includes developing deep learning models that use physiological data for recognizing psychological states and integrating the predicted states with LLMs for empathic interaction. We showcase the application of this approach in an Empathic LLM (EmLLM) chatbot for stress monitoring and control. We also discuss the results of a pilot study that evaluates this EmLLM chatbot based on its ability to accurately predict user stress, provide human-like responses, and assess the therapeutic alliance with the user. 
 
\end{abstract}


\maketitle

\section{Introduction}

In recent years, Large Language Models (LLMs) have made substantial strides in showcasing their expertise across multiple disciplines, such as computer science, medicine, and psychology~\cite{wang2023emotional}. They have demonstrated remarkable proficiency in processing complex linguistic structures and producing text that resonates with human-like coherence and relevance. While LLMs can process emotional nuances from our text inputs, they currently lack the capability to discern our mental and emotional states accurately. Enhancing LLMs' ability to interpret mental states could make them more adept in applications such as psychotherapy, which would benefit from emotional understanding.

Physiological data can provide insights into a user's mental and emotional states and could enable LLMs to engage with users more empathically. This research proposes a multifaceted approach to enhance empathic interaction with LLMs. First, we aim to construct user profiles that go beyond static preferences by incorporating users' physiological data. Second, we use deep learning to infer the user's mental and emotional state with raw physiological data. Finally, we integrate the prediction into a pre-trained LLM to develop physiology-driven Empathic LLMs (EmLLM) that are aware of the user’s mental and emotional states while interacting with them.

We present the application of our approach in an EmLLM chatbot for everyday stress monitoring and control. Users' physiological data, including Electrodermal Activity (EDA), Blood Volume Pressure (BVP), and Skin Temperature (ST), is collected using the Empatica EmbracePlus smartwatch. An end-to-end multi-channeled 1D Convolution Neural Network (CNN) trained on the Wearable Stress and Affect Detection (WESAD) dataset~\cite{schmidt2018introducing} is applied to the collected raw data to predict if a user is feeling stressed. Finally, the Falcon-7B LLM~\cite{almazrouei2023falcon} is customized to include the predicted labels to develop an EmLLM chatbot for stress monitoring and control. We discuss the findings from a pilot study with 8 students to evaluate the chatbot's ability to predict user stress, provide human-like responses, and assess therapeutic alliance.

\section{Background}

Empathy is defined as the ability to feel the same way as others (affective empathy) and understand the thoughts and perspectives of others (cognitive empathy)~\cite{gladstein1983understanding}. Humans can have varying degrees of empathy based on their backgrounds and experiences~\cite{baron2011zero}. On the other hand, LLMs are designed to understand and generate human-like text and can mimic empathy but don't actually feel emotions. However, LLMs can be trained, customized, and integrated with multimodal human data to help them understand human thoughts and emotions. Such LLMs could be useful in applications such as customer service, education and training, virtual assistants, and psychotherapy.

Psychotherapy involves conversations between a trained therapist or counselor and the individual or group seeking mental health support. With technological progress, Artificial Intelligence (AI) chatbots have emerged to provide mental health support to users. For example, Woebot, a fully automated chatbot, follows principles of Cognitive Behavior Therapy (CBT) to deliver accessible mental health support to college students~\cite{fitzpatrick2017delivering}. BioBase, another AI chatbot, combines features of wearable devices (tracking sleep, heart rate, and activity) and CBT to provide mental health support to users~\cite{ponzo2020efficacy}. However, a limitation of such chatbots is that they have a fixed conversational style for interacting with users, which can feel generic and robotic.

LLMs have emerged as transformative tools in the field of psychotherapy and offer unique opportunities to provide users with personalized mental health support. Existing research explores the potential applications and challenges of using LLMs in psychotherapy. Demszky et al.~\cite{demszky2023using} used LLMs for psychological measurement, experimentation, and practice, highlighting their ability to simulate realistic conversations for training mental health professionals. Lai et al.~\cite{lai2023supporting} proposed the Psy-LLM framework that serves as an assisting tool for mental healthcare professionals. However, there are concerns about the LLMs' ability to produce reliable outputs consistently.


Another limitation of existing AI chatbots is that users must actively engage with them to update their mental and emotional states. However, it's not always possible for users to engage with the chatbot or keep track of their mental states throughout the day. In contrast, wearable devices could automatically track the user's psychological state and passively inform the chatbot. Integrating wearable sensors with LLMs is becoming more common in biomedical applications. Belyaeva et al.~\cite{belyaeva2023multimodal} developed a multimodal LLM for healthcare that uses high-dimensional modalities, including wearable data, to estimate underlying disease risk. Liu et al.~\cite{liu2023biosignal} introduce ``BioSignal Copilot,'' a system that utilizes LLMs to interpret biomedical signals and draft clinical reports. These studies exemplify how LLMs can translate complex physiological data into actionable insights, thereby paving the way to enhance personalized psychotherapy approaches.

\section{Physiology-Driven Empathic LLMs (EmLLMs)}

To enhance empathy in LLMs and make them more suitable for applications such as psychotherapy, we propose a physiological computing approach that primarily comprises two components \cite{}:
(1) A deep learning model to predict user's psychological states with their physiological data;
(2) Integrating the predictions with a custom LLM that interacts with users based on their psychological states.
This section describes the conceptual approach of the physiology-driven EmLLM (Figure~\ref{approach}).

\subsection{Pyscho-Physiological Inference}

Human physiology and psychology are intricately linked and profoundly influence each other. The method for estimating psychological states from peripheral physiological signals is called psycho-physiological inference \cite{fairclough2009fundamentals}. Researchers extract various time-domain, frequency-domain, and statistical features to capture relevant information from the raw physiological data~\cite{schmidt2018introducing,hickey2021smart,hilty2021scoping,gedam2021review}. These features are then applied to various Machine Learning (ML) algorithms, including Decision Tree (DT), Random Forest (RF), and Support Vector Machine (SVM) to predict users' psychological states~\cite{schmidt2018introducing,gupta2020affectivelyvr}. However, there is no consensus on the best physiological features, and researchers typically use the features that give them the best accuracy.


Deep Learning (DL) networks can incorporate the feature engineering process in the modeling task and remove the dependence on hand-crafted features~\cite{ismail2019deep}. CNNs have the ability to automatically extract useful time-varying features from raw physiological data and send it to a fully connected Multilayer Perception (MLP) to perform classification~\cite{zhao2017convolutional}. CNNs can also be combined with recurrent networks, such as Long Short-Term Memory (LSTM), to develop hybrid networks for time-series classification~\cite{yao2017deepsense,du2018attention}. With CNNs, raw data sampled at varying frequencies from different sensors can be fused by applying convolutional layers to each univariate series individually and then concatenating the learned features to an MLP for classification~\cite{zheng2014time}. Finally, deep networks can also predict multiple related tasks in one network~\cite{caruana1997multitask}, thereby eliminating the need to develop different models for predicting related psychological states.

\subsection{Dynamic LLM Adaptation}


Pre-trained LLMs such as GPT-3 can perform a wide range of natural language processing tasks out of the box but require some customization for specialized applications. The common ways of customizing LLMs include techniques such as prompt engineering and model fine-tuning. Prompt engineering is a lightweight yet powerful technique for customizing the behavior of LLMs by giving them a set of instructions to perform better on particular tasks. The task performance largely depends on how the prompts are crafted, and various prompting techniques~\cite{white2023prompt} could be used to instruct an LLM to get the desired output.  Prompt engineering is a promising technique for integrating psycho-physiological inferences with LLMs to make them more empathetic. Another approach for customizing LLMs is model fine-tuning, which refers to the process of updating the parameters of a pretrained LLM on a task-specific dataset. There is little scope for integrating psycho-physiological inferences with LLM fine-tuning, but it is a powerful approach to control LLMs' output.

To develop the EmLLMs, the predicted user psychological states from the deep learning model can be used as part of the crafted prompts for the LLM. These prompts can also instruct the LLM to adopt an empathetic personality~\cite{white2023prompt} and follow chain-of-empathy reasoning in conversations to understand what the user is going through~\cite{lee2023chain}. LLMs can also be instructed to follow the principles of various psychotherapy techniques, such as CBT and Active Listening, to enhance their empathy. Moreover, fine-tuning the LLMs on specialized datasets can generate more context-specific responses, enhancing the user experience. LLMs can also be trained from scratch on specialized datasets~\cite{lai2023supporting}, but this approach requires more time and resources. Lastly, LLMs can go beyond natural language comprehension and can be trained on multimodal data to power conversation around user health and wellness~\cite{belyaeva2023multimodal}.

\section{Physiology-Driven EmLLM Prototype}

Based on the conceptual framework of EmLLM, we developed a prototype chatbot that passively monitors user stress throughout the day using their physiological data and actively interacts with them at the end of the day based on how much stress they experienced. We use the Empatica EmbracePlus smartwatch~\cite{CiteDrive2022_1} to collect users' raw physiological data, including BVP, EDA, and ST, which is used as input to an end-to-end multi-channeled 1D CNN model.

The model inputs the raw physiological data individually through separate channels and applies a sequence of three 1D convolution layers with a kernel size of 3 and a stride length proportional to the sampling rate of the physiological signals. The convoluted features from each channel are then pooled (1D max-pooling), flattened, and concatenated to pass it over to fully connected layers for classification. We use the ``Adam'' optimizer and ``Binary Cross Entropy'' loss function to train our model on the WESAD dataset~\cite{schmidt2018introducing} to predict if a participant is feeling ``stressed'' or ``not stressed''. 
We followed an approach similar to Schmidt et al.~\cite{schmidt2018introducing} for binary stress classification and combined all participants' data from baseline and amusement conditions to a non-stress class. The authors used a variety of hand-crafted features to train several ML models, and the maximum \textit{accuracy and F1 score} with the wrist data from all participants was \textit{84.1\%} and \textit{86.9 F1 score}. However, our model achieved a state-of-the-art \textit{accuracy of 85.1\%} and \textit{89.0 F1 score}.

To develop the EmLLM chatbot for stress-related psychotherapy, we start with fine-tuning the Falcon-7B LLM~\cite{almazrouei2023falcon} using the Quantized Low-Rank Adaption (QLoRA) technique~\cite{dettmers2024qlora} on the mental health conversational dataset~\cite{CiteDrive2022}. The Falcon-7B LLM was selected because it is open source and has extensive adaptability. The QLoRA technique involves normalizing and reducing the model's weights to a 4-bit quantized format, significantly decreasing the memory requirement. The mental health dataset comprises a rich collection of conversational pairs, including patient questions and corresponding answers provided by mental health professionals. Next, the fine-tuned LLM is prompt-engineered to take as input the predicted user stress levels, act like a trained psychologist, follow the principles of CBT, and provide a reasonable explanation whenever it cannot answer a question. Lastly, we develop a web user interface for the EmLLM chatbot for psychotherapy.

\section{Pilot Study}

\subsection{Study Protocol}

The study aimed to evaluate the chatbot's ability to accurately predict user stress, provide human-like responses, and assess the therapeutic support provided to the user.
To achieve this, we conducted an in-the-wild study to monitor the user's physiology in their work environment throughout the day and make them interact with the chatbot at the end of the day. The different parts of the study protocol are detailed below- 

\noindent \textbf{Preparation:}
Participants arrive at the study location before they begin their daily tasks. First, they complete the pre-task Short Stress State Questionnaire (SSSQ)~\cite{helton2015short}. Then, the participants are equipped with the Empatica EmbracePlus smartwatch.

\noindent \textbf{Data Collection:}
After completing the preparation phase, the participants were asked to resume their daily tasks. However, they were instructed to keep wearing the smartwatch for the entire study duration. Participants were advised not to loosen or change the position of the smartwatch. They are asked to avoid any mood-altering substances, including coffee and tobacco, and not to do any strenuous exercise during the study period.

\noindent \textbf{Interaction:}
At the end of the day, participants returned to the study location. First, the smartwatch was removed from the participants' wrists, and then they were asked to complete the post-task SSSQ. Finally, the participants were asked to interact with the EmLLM chatbot for at least 15 minutes. The chatbot greets the participant by their name and introduces itself. It then informs the participant about their predicted stress and asks if they want to discuss their day with it. After completing the session, the participants were asked to complete the Godspeed questionnaire~\cite{bartneck2023godspeed} and the Session Rating Scale (SRS)~\cite{duncan2003session}. More details about the survey instruments are given in the section below. We also asked participants to comment or give feedback.

\subsection{Survey Instruments}

The survey instruments used in this study, SSSQ, Godspeed questionnaire, and SRS, assess the chatbot's ability to predict user stress, provide human-like responses, and give therapeutic support to the user, respectively. The SSSQ is an adaptation of the Dundee Stress State Questionnaire (DSSSQ), which measures three higher-order dimensions of subjective stress states: engagement, distress, and worry~\cite{helton2015short}. The pre-and-post-SSSQ provides a reflection of the stress states that the user experienced during the day while performing their daily tasks. We use the results of the pre-and-post-SSSQ to validate the stress predictions.

The Godspeed questionnaire is a tool used in human-robot interaction research to gather users' feedback about their experiences interacting with robots~\cite{bartneck2023godspeed}.
We use its answers to assess the user's perception of the chatbot's friendliness, perceived intelligence, ease of use, and likability instantly after the counseling session.
The SRS is a tool commonly used in psychotherapy to assess clients' subjective experiences of individual therapy sessions~\cite{duncan2003session}. We use its responses to gather user feedback immediately after each session to assess the therapeutic alliance between the user and the chatbot.

\subsection{Results}

We used the study protocol to evaluate the performance of the EmLLM chatbot for psychotherapy among 8 students. Among these, 5 identified themselves as men and 4 as women, ages 23 to 37 (mean age = 30.5). For our pilot study, we wanted to focus on a population that experiences high degrees of work-related mental stress. Therefore, we focus on Ph.D. students as they experience high-stress levels resulting from their academic pursuits~\cite{satinsky2021systematic,roofigari2023conceptual}.

\begin{figure*}[!t]
    \centering
    \begin{minipage}[t]{0.49\textwidth}
        \centering
        \includegraphics[width=\textwidth]{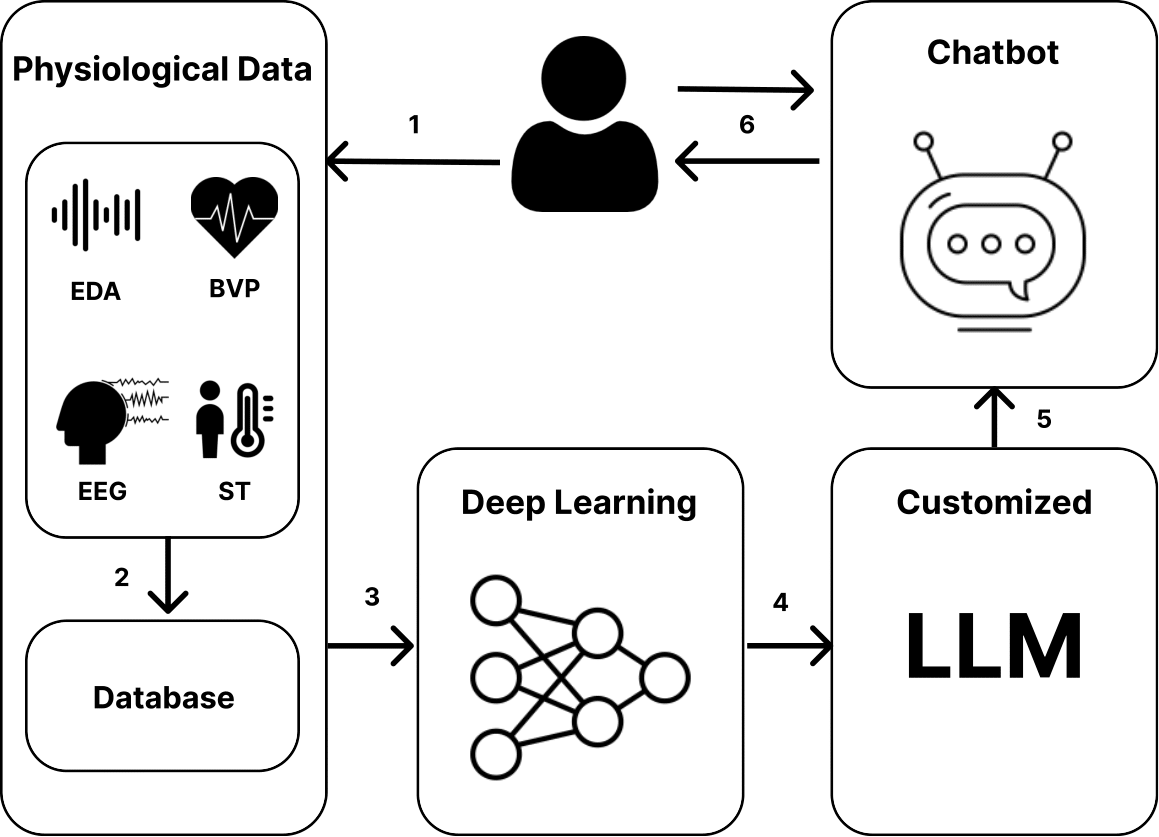}
        \caption{Proposed EmLLM approach.}
        \label{approach}
    \end{minipage}
    \hfill
    \centering
    \begin{minipage}[t]{0.49\textwidth}
        \centering
        \includegraphics[width=\textwidth]{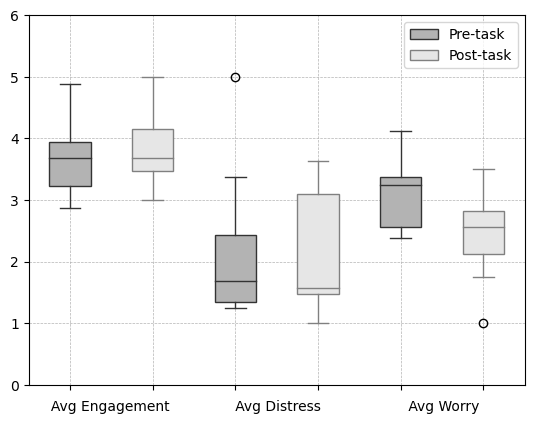}
        \caption{Average stress states of all participants.}
        \label{results}
    \end{minipage}
\end{figure*}

\subsubsection{Quantitative Results}

Figure~\ref{results} shows the average values of the three stress states for all participants before and after they performed their daily work-related tasks. As expected, almost all participants experienced high levels of worry and engagement before and after performing their daily tasks. The post-task worry was less than the pre-task, but participants felt more engaged later in the day, suggesting an inability to detach from work psychologically, indicating more stress~\cite{sonnentag2005switching}. Some participants also experienced higher degrees of distress after their workday ended.

The results of the stress prediction model from physiological data show that 6 out of 8 participants experienced some stressful events in their workday. The predictions are consistent with the findings of the SSSQ, indicating that the stress prediction model is more likely to predict the stress state of engagement. The two participants with no signs of stress in the prediction model also showed high levels of engagement, with one scoring high in all three stress states in both pre-task and post-task SSSQ. These participants, however, belong to special groups, with one having medically diagnosed Attention-Deficit/Hyperactivity Disorder (ADHD) and the other with a hearing disability.

Chatbot's mean competence score was 3.5, and its mean responsiveness score was 3.63.  However, it appears that participants were not very impressed by its ability to give human-like responses (2.88), conversational elegance (2.88), and pleasantness (3). The chatbot also seemed to have developed a good therapeutic alliance with participants, with a mean quality rating of 3.25 and a mean empathy score of 3.63. The participants also felt that the psychotherapy session with the chatbot was relevant (3.75), a good fit (3.38), and right (3.25) for them.

\subsubsection{Qualitative Results}

Along with the survey responses, the participants also gave their comments and feedback on their interaction with the chatbot. Most participants thought the EmLLM chatbot was ``therapist-like'' and gave ``excellent counsel.'' They were impressed that the chatbot ``asked questions to make them think'' and ``gave context'' for asking these questions instead of giving generic information like ChatGPT. However, a common concern raised by all participants was that sometimes the chatbot would ``start talking in Spanish''. This could be because the mental health dataset used to fine-tune the LLM had some conversation pairs in Spanish. Some participants were also surprised by the stress prediction made by the chatbot about their day, as it was opposite to their expectations. Participants also wanted the chatbot to remember their conversation and give them more details about their day from the smartwatch data. However, one participant was concerned about their chat history and thought ``someone would see'' their conversation with the chatbot.

\section{Discussion}


\noindent \textbf{Accuracy of Psycho-Physiological Inference}
The stress prediction model in this study combines the physiological data of all participants subjected to baseline, amusement, and stressful conditions. The data from baseline and amusement conditions are further combined to make stress prediction a binary classification task. Combining the data does give a stress prediction model of higher accuracy, but it raises concerns about the model's diagnosticity, sensitivity, reliability, and validity. Diagnosticity and sensitivity refer to the precision of inference, reliability captures the generality of inference across people, and validity represents the consistency of inference across conditions \cite{fairclough2017physiological}. HCI researchers must investigate these topics when using psycho-physiological data in their studies.

\noindent \textbf{Does more data necessarily mean more accuracy?}
ML research for predicting human psychological states such as stress, emotion, and thermal comfort with wearable data uses both physiological and non-physiological data. Physiological data such as BVP, EDA, and ST are scientifically proven to be related to human psychology. Non-physiological data collected from wearable devices, including ACC, is useful in predicting human activity but has little relation to the mental states of humans. Data-hungry ML models will give better accuracy with more data and features, but including non-physiological data to infer psychological states is fundamentally questionable and can result in models that overfit.

\noindent \textbf{Are hand-crafted features indeed better?}
Another practice that potentially overfits psychophysiological ML models is using several hand-crafted features. Researchers might use techniques such as correlation analysis, principal component analysis, and regularization to remove redundant features and avoid overfitting. However, there is no consensus on the best features; researchers typically use features that give them the best accuracy. A possible solution can be using end-to-end deep learning models incorporating the feature extraction process from raw data in the classification pipeline without relying on hand-crafted features. In this study, we developed an end-to-end multichannel 1D CNN that gave us state-of-the-art results, but this model can be further improved by combining it with recurrent networks.


\noindent \textbf{Biocybernetic Adaptations in LLMs}
Biocybernetic adaptation is a category of physiological computing systems that use physiological data to infer users' psychological states and adapt based on these states to improve user experience \cite{fairclough2014meaningful}. This research integrates psycho-physiological inferences with LLMs using prompt engineering to facilitate empathic human-AI interactions. However, the behavior of LLMs is complex, and prompt engineering alone cannot ensure the necessary adaptation in LLMs. Training and fine-tuning LLMs on specialized multi-modal datasets can help control their output, but this requires adapting the model's input layers and learning algorithms to process and correlate data across different sensory modalities effectively.

\noindent \textbf{Privacy, Security, and Ethics}
Developing technological solutions for applications such as mental health support introduces substantial privacy, security, and ethical challenges for users \cite{jones2023privacy}. Inadequate management of psycho-physiological data and users' intimate conversations with the chatbot could lead to breaches of confidentiality and unintentional disclosures. Implementing robust data protection measures and stringent ethical guidelines is crucial to ensure the privacy and security of sensitive information. Such precautions are essential for maintaining user trust and safeguarding the well-being of individuals who rely on these digital interventions.





\section{Conclusion}

Our research advances empathic AI by integrating Large Language Models (LLMs) with physiological data, creating the EmLLM chatbot for stress monitoring and control. This integration enables the chatbot to interpret users' psychological states more accurately, which could enhance the effectiveness of digital psychotherapy tools. The pilot study validates our approach, showing improved stress prediction and the ability to build a therapeutic rapport. Future work should focus on developing AI that can provide personalized mental health support by further understanding individual physiological responses, aiming to make digital therapy more effective and responsive to users' needs.

\bibliography{sample-ceur}




\end{document}